\documentclass[fleqn,usenatbib]{mnras}
\usepackage{newtxtext,newtxmath}
\usepackage[T1]{fontenc}
\usepackage{graphicx}	
\usepackage{amsmath}	
\usepackage{mathtools}                                             
\usepackage{subfigure}  
\usepackage{color}                                                 
\usepackage{xcolor}                                             
\usepackage{hyperref}   
\usepackage{placeins}

\date{}
\pubyear{2022}
\title[The Abell~399-401 pair]{Radio multifrequency observations of the galaxy cluster pair Abell~399$-$401 with WSRT}

\author[C.~D.~Nunhokee et al.]
{C.~D.~Nunhokee$^{1,2,3}$\thanks{email: ridhima.nunhokee@curtin.edu.au}, G.~Bernardi$^{4,5,6}$,
	S.~Manti$^7$, F.~Govoni$^8$, A.~Bonafede$^{4,9}$, T.~Venturi$^4$, \newauthor  D.~Dallacasa$^{4,9}$, M.~Murgia$^8$, R.F.~Pizzo$^{10}$, O.M.~Smirnov$^{5,6}$ and V.~Vacca$^8$\\
	\\
	$^1$International Centre for Radio Astronomy Research, Curtin University, Bentley, WA 6102, Australia\\
	$^2$ARC Centre of Excellence for All Sky Astrophysics in 3 Dimensions (ASTRO 3D), Bentley, WA 6102, Australia \\
	$^3$Department of Astronomy, University of California, Berkeley, CA\\
	$^4$INAF - Istituto di Radioastronomia, via Gobetti 101, 40129 Bologna, Italy\\
	$^5$Department of Physics and Electronics, Rhodes University, PO Box 94, Grahamstown, 6140, South Africa\\
	$^6$South African Radio Astronomy Observatory, Black River Park, 2 Fir Street, Observatory, Cape Town, 7925, South Africa\\
	$^7$Scuola Normale Superiore, Piazza dei Cavalieri 7, 56126, Pisa, Italy\\
	$^8$INAF - Osservatorio Astronomico di Cagliari Via della Scienza 5, I-09047 Selargius (CA), Italy\\ 
	$^9$Dipartimento di Fisica e Astronomia, Universit\'{a} di Bologna, Via Gobetti 93/2, 40129 Bologna, Italy\\
	$^{10}$ASTRON, Netherlands Institute for Radio Astronomy, PO Box 2, 7990 AA Dwingeloo, The Netherlands}

\begin{document}
	\label{firstpage}
	\pagerange{\pageref{firstpage}--\pageref{lastpage}}
	\maketitle
	
	\begin{abstract} 
		Galaxy clusters are assembled via merging of smaller structures, in a process that generates shocks and turbulence in the intra cluster medium and produces radio diffuse emission in the form of halos and relics. The cluster pair A\,$399-$A\,401 represents a special case: both clusters host a radio halo. Recent Low Frequency Array (LOFAR) observations at 140~MHz revealed the presence of a radio bridge connecting the two clusters along with two relic candidates. These relics include one South of A\,399 and the other in between the two clusters, in proximity of a shock front detected in X-ray observations. In this paper we present observations of the A\,$399-$A\,401 cluster pair at 1.7, 1.4, 1.2~GHz and 346~MHz from the Westerbork Synthesis Radio Telescope (WSRT). We detect the radio halo in the A\,399 cluster at 346~MHz, extending up to $\sim 650$~kpc and with a $125 \pm 6$~mJy flux density. 
		Its spectral index between 140~MHz and 346~MHz is $\alpha = 1.75 \pm 0.14$.
		The two candidate relics are also seen at 346~MHz and we determine their spectral indices to be $\alpha = 1.10 \pm 0.14$ and $\alpha = 1.46 \pm 0.14$. 
		The low surface brightness bridge connecting the two clusters is below the noise level at 346~MHz, therefore we constrain the bridge average spectral index to be steep, i.e.  $\alpha > 1.5$ at 95\% confidence level. This result favours the scenario where dynamically-induced turbulence is a viable mechanism to reaccelerate a population of mildly relativistic particles and amplify magnetic fields on scales of a few Mpcs.
	\end{abstract}
	

	\section{Introduction}
	\label{sec:introduction}
	Clusters of galaxies are located at the nodes of the cosmic web and formed by subsequent merging of smaller structures (e.g.~groups or clusters of galaxies). These merger events likely accelerate particles and amplify magnetic fields in the intra-cluster medium that generate diffuse radio sources 
	in the form of radio halos and relics \citep[see][for recent reviews on the topic]{Brunetti2014,vanWeeren2019}. Radio halos and relics are observed at the centre and periphery of galaxy clusters, respectively, through diffuse synchrotron emission. Although it is largely accepted that radio halos are generated by merger-induced turbulence \citep[e.g.,][]{Cassano2005,Brunetti16a,Pinzke17} and radio relics are created by shocks \citep[e.g.,][]{Hoeft07,Kang16}, several questions on the acceleration efficiency and, therefore, the details of the re-acceleration mechanism remain still open \citep[e.g.,][]{Brunetti16b,Wittor17}.
	
	The A\,$399-$A\,401 pair has a special place in the cluster landscape.
	It is a local pair \citep[$z = 0.0718$ and $z = 0.0737$, respectively,][]{Oegerle2001}, 
	separated by a projected distance of $\sim 3$~Mpc. The masses of the A\,$399-$A\,401 pair are $5.7 \times 10^{14}$~M$_\odot$ and $9.3 \times 10^{14}$~M$_\odot$, with similar gas temperatures of $kT \approx 7$ and 8~keV respectively \citep{Fujita1996, Fabian1997, Markevitch1998}. It has a filament of X-ray emitting gas in the interconnecting region \citep{Akamatsu17}. Both X-ray and optical observations \citep{Bonjean18} indicate that the system is likely in the early phase of a merging event. Observations of the Sunyaev-Zeldovich effect \citep{Bonjean18,Hincks22} confirmed the presence of a connecting bridge between the two clusters, 
	with a $4.3 \times 10^{-4}$~cm$^{-3}$ gas density \citep{Bonjean18}.
	
	At radio wavelengths, both clusters host a radio halo whose integrated flux density at 1.4~GHz is 
	$S = 20.4$~mJy and $S = 19.3$~mJy respectively \citep{Murgia2010,Govoni2019}. The A\,399 halo has a typical roundish morphology and extends to $\sim 570$~kpc, whereas the A\,401 halo has a more irregular morphology and extends up to $\sim 350$~kpc.  
	This pair was the target of recent, deep LOFAR observations at 140~MHz \citep{Govoni2019} that detected both radio halos, extending up to 970~kpc and 800~kpc respectively. These observations also revealed a series of diffuse emission features that are not visible at GHz frequencies, in particular, a bridge of radio emission connecting the two halos.
	They provide the first evidence of relativistic particles and magnetic fields between the two galaxy clusters, at distances comparable to the cluster virial radius \citep[together with the A\,1758 double system;][]{Botteon18,Botteon2020}.
	Radio emission on such large scales in the bridge may be due to a distribution of weak shocks that fill the bridge volume and reaccelerate a pre--existing population of mildly relativistic particles \citep{Govoni2019}.
	
	Conversely, \citet{Brunetti2020} propose an alternative scenario where acceleration happens via second--order Fermi mechanisms: relativistic particles scatter with magnetic field lines diffusing in super-Alfvenic turbulence which also amplifies magnetic fields. In this case, steep-spectrum synchrotron emission can be generated in the entire intra-cluster bridge region.
	Spectral index measurements of the bridge emission are therefore necessary to understand the particle acceleration mechanism in action here.

	In this paper we present observations at 1.7, 1.4, 1.2~GHz and 346~MHz (18, 21, 25 and 92~cm respectively) of the A\,399-A\,401 cluster pair, aimed to characterize the 
	properties of their diffuse radio emission.
	The paper is organized as follows: observations are described in Section~\ref{sec:observations}, and the data reduction and calibration are discussed in Section~\ref{sec:calibration}. Results are presented in Section~\ref{sec:results} and 
	conclusions are offered in Section~\ref{sec:conclusions}. 
	Throughout the paper we used Planck cosmology, where $1'' = 1.345$~kpc at the distance of the cluster pair, the Hubble parameter $h=0.72$, the total matter density $\Omega_M=0.253$ and the cosmological constant $\Omega_{\Lambda}=0.742$ \citep{Planck2020}.


\section{Observations}
\label{sec:observations}
Observations were carried out with the WSRT at four frequency bands centred at 1.7, 1.4, 1.2~GHz and 346~MHz. 
The WSRT is an aperture synthesis array composed of 14 dishes with a diameter of 25~m, arranged on an east-west track, that uses the Earth rotation to fill the $uv$~plane in a 12~hour synthesis observation. 
Ten telescopes have a fixed location, and are spaced by a distance of 144~m whereas the four remaining dishes (identified with the A, B, C and D letters respectively) can be moved along two rail tracks to provide different array configurations.

\begin{figure}
	\includegraphics[width=1.\columnwidth,angle=0]{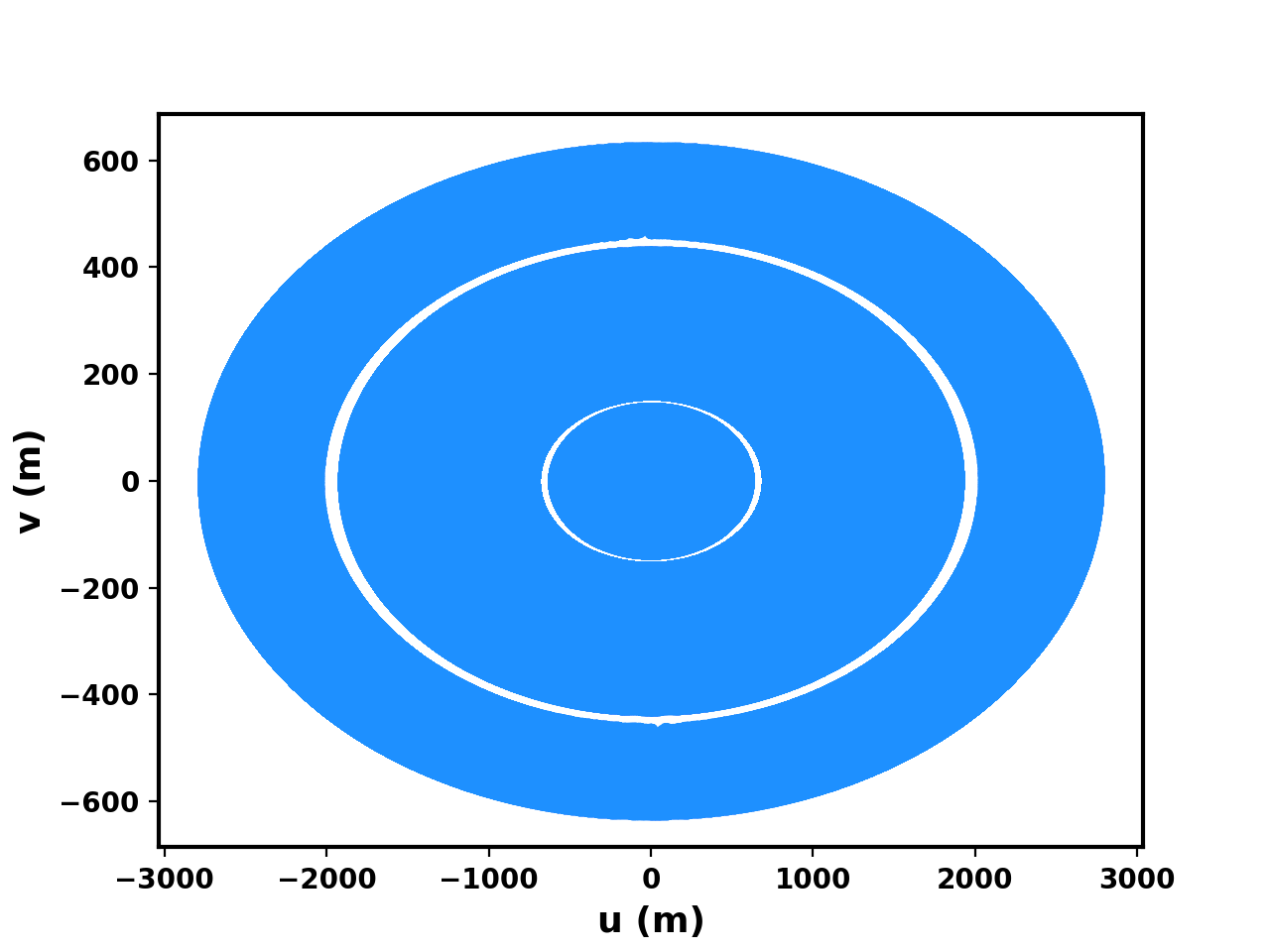}
	\caption{Monochromatic $uv$ coverage at 346~MHz, including all the six configurations. The two gaps are due to antenna~5 that was missing from all observations. The low declination of the cluster pair results in a fairly ellipsoidal $uv$ coverage, which, in turn, results in a limited resolution along declination (Table~\ref{tab:table_obs}).}
	\label{fig:uvcov_92cm}
\end{figure}

Observations were split into two night periods: the 1.7 and 1.4~GHz bands were conducted in December 2010 and
the 1.2~GHz and 346~MHz bands were taken in November 2011. The telescope was pointed at the centre of the A\,399 cluster.
The 1.7 and 1.4~GHz observations were carried out with the so--called Maxi-Short configuration\footnote{see \url{https://www.astron.nl/radio-observatory/astronomers/wsrt-guide-observations/3-telescope-parameters-and-array-configuration} for a description of the configurations in terms of antenna separations}, with  the four shortest baselines being 36, 54, 72, and 90~m. This configuration maintained regular $uv$-coverage, aimed to provide good sensitivity to extended structures with dense $uv$~coverage between 36 and 2760~m baselines. The 1.2~GHz measurements were taken with four different configurations, where the four movable telescopes were moved in 18~m increments, from 36 to 90~m. Whilst the 346~MHz observations used 
six configurations, with the four movable telescopes stepped at 12~m increments and the shortest spacing ran from 36 to 96~m, pushing the grating lobe out to a radius of $\sim 1^\circ$. This provided $uv$~coverage with  baselines ranging from 36 to 2760~m as shown in Figure~\ref{fig:uvcov_92cm}.  The observing band is divided into eight contiguous sub bands for a total of 80~MHz bandwidth at 92~cm and 150~MHz at each of the higher frequencies. Noise levels are calculated at the edge of the field of view where the primary beam attenuates the intrinsic sky emission. 
Each observing run included a $\sim$ 20~minute observation of a calibration source. 
Details of the observing setup are presented in Table~\ref{tab:table_obs}.

\begin{table*}
	\caption{Setup for WSRT observations centred at 1.7, 1.4, 1.2~GHz and 346~MHz.}
	\centering
	\begin{tabular}{@{}ccccccccc}
		\hline
		\hline
		Reference  & Frequency 	& Channel 	& Integration & Pointing  & Synthesized   &   Noise  & Calibrator & Calibrator \\
		wavelength & range          &    width    & time            & centre     & beam               & rms       &  source                & Duration\\
		(cm)		     & (GHz)		&  (MHz)      & (hours)       & (degrees)   &(arcsec)       &  (mJy~beam$^{-1}$)	& (346~MHz)    & (minutes)\\
		\hline
		18 		     & $1.64-1.79$	&  1.17 &   12  & (44.5, 13.01)  & $53 \times 9$   &   0.078	&  3C348 &	20	\\
		21		     & $1.30-1.46$	&  1.17	&   12  &   (44.5, 13.01) & $53 \times 11$  &   0.059 		& 3C348 & 	20\\
		25		     & $1.15-1.29$	&  1.17 &    $4 \times 12$ &  (44.5, 13.01) &  $58 \times 12$  &   0.063 & 3C348  &	20	\\
		92		     & $0.31-0.38$	&  0.156  & $6 \times 12$  & (44.5, 13.01) & $205 \times 43$ &   1.2	 &  3C295 &	20 \\
		\hline
	\end{tabular}
	\label{tab:table_obs}
\end{table*}

\section{Data Reduction and Calibration}
\label{sec:calibration}
Data were initially tapered using a Hanning window. The edges of each sub-band were discarded. Radio Frequency Interference (RFI) were identified and flagged using the AOFlagger package \citep{Offringa2010}. Data were further reduced using the Common Astronomy Software Applications (CASA)\footnote{http://casa.nrao.edu} software and integrated with routines specifically developed for calibration of WSRT data \citep{Bernardi2009, Bernardi2010}. The absolute flux calibration was set to the \citet{Scaife2012} scale using observations of 3C295 at 346~MHz.

	The reduction of the GHz data (1.7, 1.4 and 1.2) can be summarized by the following steps. An initial bandpass calibration was derived from observations of 3C48 and applied to the target field. A dirty image was generated by Fourier transforming the visibilities. All the sub bands were combined together using the multifrequency synthesis algorithm with uniform weights. The dirty image was deconvolved using the Cotton--Schwab algorithm down to a threshold of 0.1~mJy~beam$^{-1}$, corresponding to the first negative component in the model. The model was then used for self--calibration where antenna based phase solutions were computed every minute. The self--calibration solutions were applied to the data and further flagging was performed on the residual visibilities. The resulting visibilities were finally imaged and deconvolved down to thresholds of 80~$\mu$Jy~beam$^{-1}$ for 1.7~GHz and 0.15~mJy~beam$^{-1}$ for 1.4 and 1.2~GHz.

At 346~MHz, the initial bandpass calibration was derived from observations of 3C295 for each night and applied to the data. All the six configurations and eight sub bands were imaged jointly using the multifrequency synthesis algorithm. The subsequent deconvolution and self--calibration followed the same path as the 1.7~GHz band reduction but with phase solutions computed every 10~seconds. The final image was obtained after deconvolving down to a 3~mJy~beam$^{-1}$ threshold.

The system noise between the calibrator and the target field can vary significantly at low frequencies and cannot be safely calibrated using the WSRT online loop gain system due to RFI contamination. We estimated these variations by calculating the total power ratio between 3C295 and the target field in clean areas of the spectrum \citep[e.g.,][]{Brentjens2008, Pizzo2009, Bernardi2009}. We found the ratio to be $\sim 5\%$ averaged over the 346~MHz band and the visibility data were corrected accordingly.


\section{Results}
\label{sec:results}
\begin{figure*}
	\centering
	\includegraphics[width=1\textwidth]{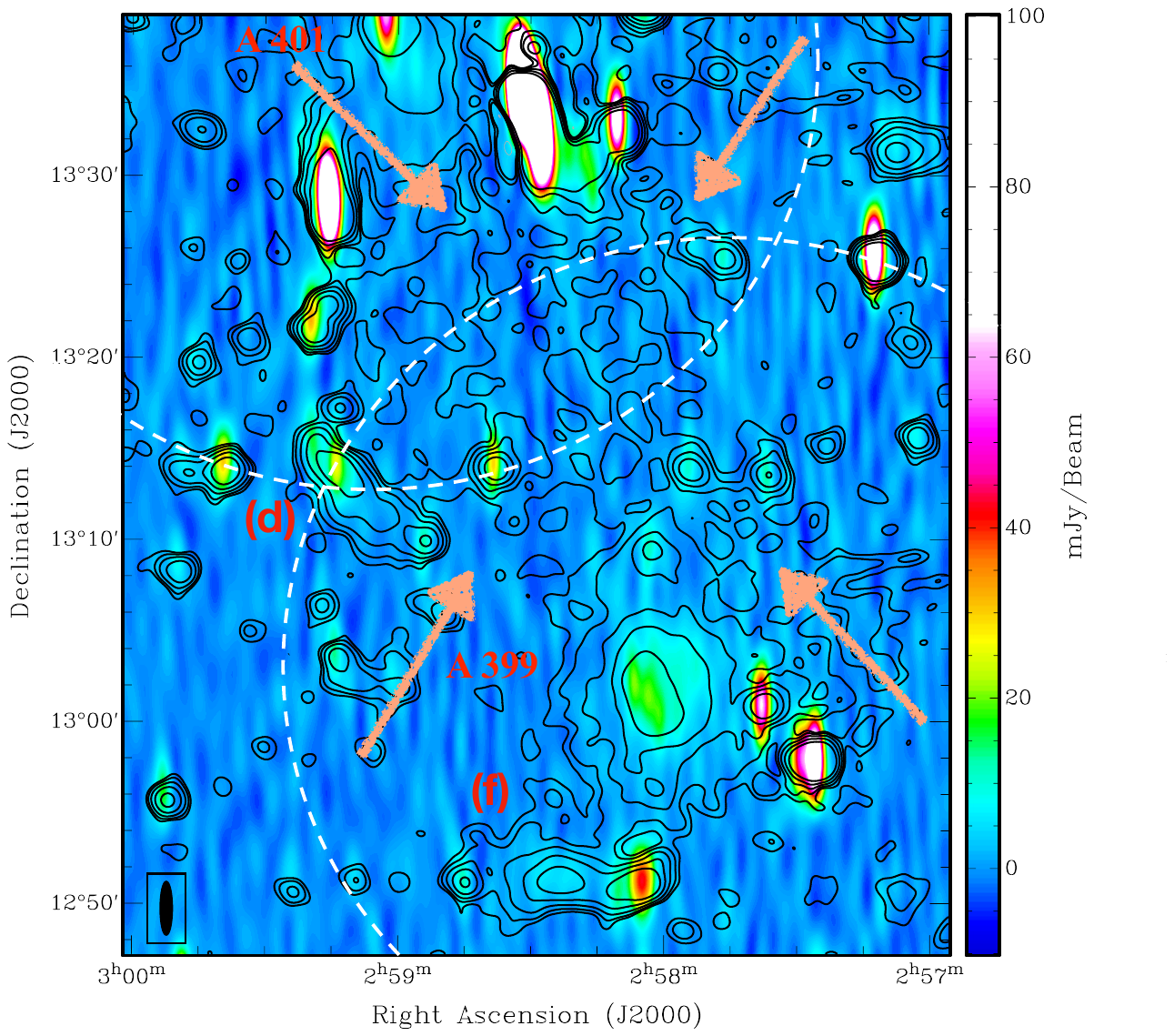}
	\caption{The 346~MHz WSRT image overlaid with 140~MHz contours \citep{Govoni2019}. The first contour is drawn at 1~mJy~beam$^{-1}$, increasing by a factor of two.  Other than the two clusters A\,399 and A\,401, labels indicate diffuse sources $d$ and $f$, identified at 140~MHz \citep{Govoni2019}. Arrows indicate the synchrotron bridge connecting both clusters. White dashed circles indicate the 1.5~Mpc virial radius, similar for both clusters \citep{Sakelliou2004}. The 346~MHz observations are pointed at the centre of the A\,399 cluster. The resolution of the 140~MHz image is $72'' \times 72''$. The beam size for the 346~MHz observations ( $205''\times 43''$)  is shown in the inset. Note that the image presented is not corrected for the primary beam attenuation. }
	\label{fig:radio_ridge}
\end{figure*}

\begin{figure}
	\centering
	\includegraphics[width=1\columnwidth]{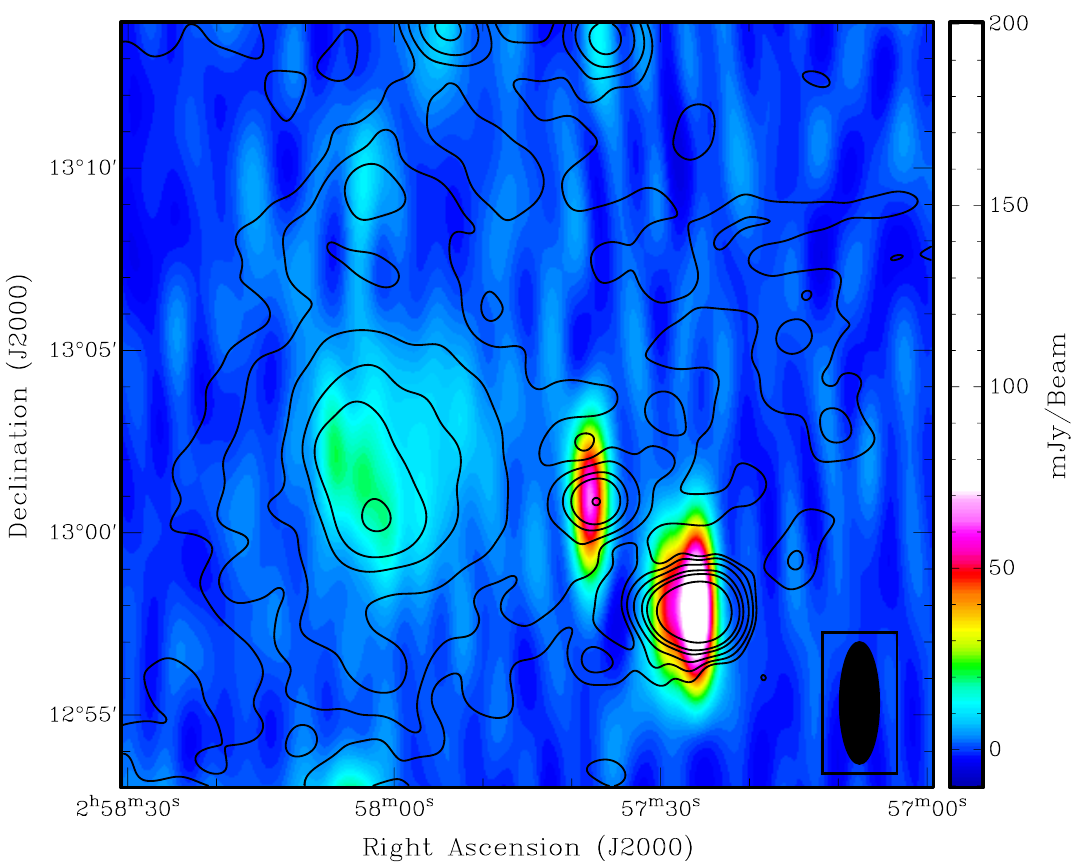}
	\caption{The 346~MHz image, with a synthesized beam of $205''\times 43''$,  overlaid with 140~MHz contours of the A\,399 radio halo \citep{Govoni2019}. The first contour is drawn at 1.5~mJy~beam$^{-1}$, with other contours spaced by a factor of two. The halo at 140~MHz extends further than at 346~MHz. The beam size for the 346~MHz observations is shown in the inset.}
	\label{fig:halo_140MHz}
\end{figure}

Our $1.2-1.7$~GHz observations are not deeper than  \citet{Murgia2010} in terms of brightness sensitivity
and therefore, do not detect the A\,399 halo (refer to Appendix~\ref{appendix:GHzimages} for images). Conversely, the halo is clearly visible in the 346~MHz image.
The size of the A\,399 halo at 346~MHz is $\sim 650$~kpc, somewhat more extended than at 1.4~GHz ($\sim 570$~kpc) and its flux density (integrated above the $4 \sigma$ contour) is $125 \pm 6$~mJy. The contour level was selected meet the level used by \citep{Govoni2019}.

There is no clear detection of the A\,401 halo neither at GHz frequencies nor at 346~MHz. The pointing was centred on A\,399, i.e. 45' away from A\,401, a distance that corresponds approximately to the first primary beam null at GHz frequencies, implying a significant attenuation of the sky emission. The radio emission observed at 346~MHz is likely the blend of two compact sources at (RA$_{\rm J2000} = 2^{\rm h} 59^{\rm m} 2^{\rm s}$, DEC$_{\rm J2000} = 13^\circ 39' 14'')$ and (RA$_{\rm J2000} = 2^{\rm h} 59^{\rm m} 1^{\rm s}$, DEC$_{\rm J2000} = 13^\circ 37' 41'')$ respectively. There is no clear emission where the 1.4~GHz halo is detected by \cite{Murgia2010} - likely due to the sensitivity loss away from the pointing centre.

Figure~\ref{fig:radio_ridge} shows the $1^\circ$-wide contours at 140~MHz \citep{Govoni2019} overlaid on the 346~MHz image. A zoomed version into the A\,399 radio halo is presented in Figure~\ref{fig:halo_140MHz}. The halo appears less extended and with a somewhat less regular morphology at 346~MHz compared to 140~MHz. The difference in morphologies may be attributed to the variation in the synthesized beams. The peak of its brightness distribution is offset by $\sim 2$~arcmin with respect to the 140~MHz observations.

\subsection{Spectral Analysis of the A\,399 Radio Halo}
\label{subsec:A399_halo}

\begin{figure*}
	\centering
	\includegraphics[width=0.98\columnwidth]{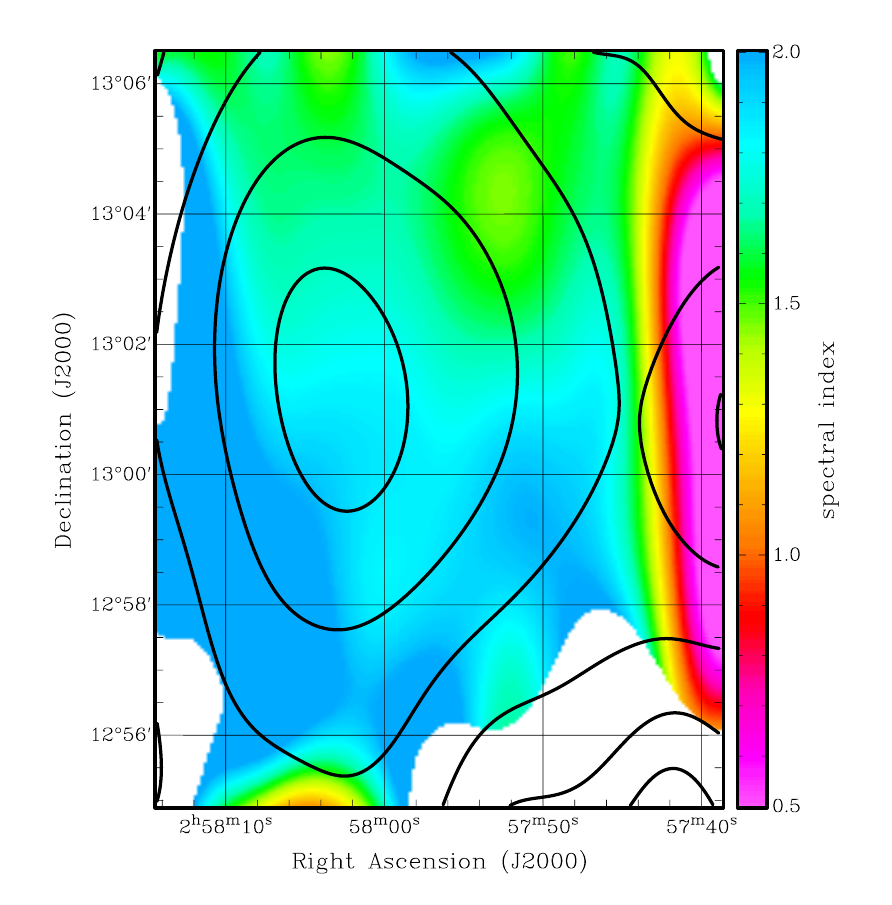}
	\includegraphics[width=0.98\columnwidth]{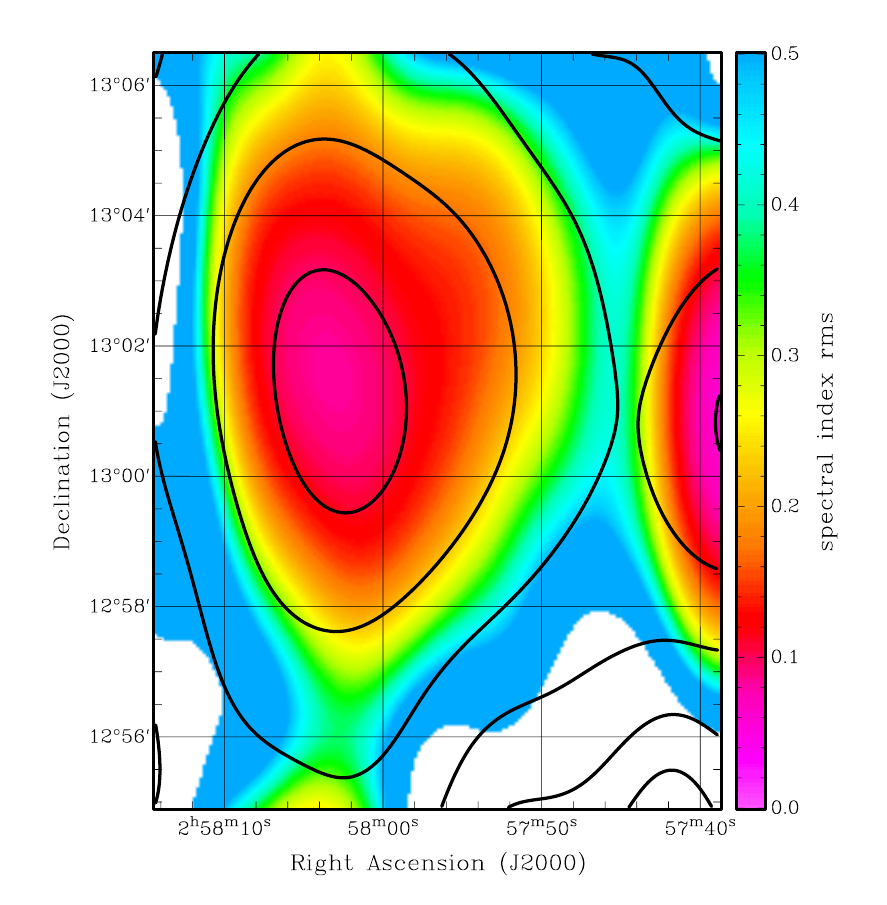}
	\caption{Left: The radio spectral index map evaluated using 140~MHz and 346~MHz observations smoothed to the same angular resolution of $205'' \times 72''$, overlaid with 140~MHz contours \citep{Govoni2019}. Right: The radio spectral index error map, overlaid with 140~MHz contours. In both panels, the first contours are drawn at 5~mJy~beam$^{-1}$ and then spaced by a factor of two.}
	\label{fig:alphaA399_140}
\end{figure*}

Given that the 346~MHz observations have an essentially complete $uv$-coverage up to $1^\circ$ scale, we derived a spectral index\footnote{We used the notation $S_\nu \propto \nu^{-\alpha}$, where $S_\nu$ is the flux density at the frequency $\nu$.} map of the A\,399 halo between 
346~MHz and 140~MHz. 
For this purpose, we used the 346~MHz map generated from Section~{\ref{sec:calibration}} and the 140~MHz images from \citet{Govoni2019}. The 140~MHz was imaged using baselines greater than 80$\lambda$ and a briggs weighting scheme (robustness of 0.25). The imaging parameters were set to achieve a consensus between sensitivity and angular resolution to be able to see the radio bridge. We applied the same logic pertinent to our 346~MHz observations. This in turn, led to a variation in the imaging parameters used by the two observations. 
	In order to match the resolution of the LOFAR observations, the 346~MHz images were smoothed to $205'' \times 72''$. The spectral index of A\,399 halo uncertainty maps $\Delta \alpha$ is calculated using:

\begin{equation}\label{eq:alpha_uncertainty}
	\Delta \alpha (x,y) = \sqrt{ \left ( \frac{\sigma_{\nu_1}}{S_{\nu_1} (x,y)} \right )^2 + \left ( \frac{\sigma_{\nu_2}}{S_{\nu_2} (x,y)} \right )^2} \ln{\frac{\nu_2}{\nu_1}}
\end{equation} 
where $(x,y)$ indicate the pixel of the map, $\nu_1 =$~346~MHz and $\nu_2 =$~140~MHz, 
 and $\sigma$ is the uncertainty associated with each pixel. The uncertainty is calculated using the quadrature sum of the image noise map and, the uncertainty on absolute flux calibration, assumed to be 5\% and 15\% at 346~MHz (refer to Section~\ref{sec:calibration}) and 140~MHz \citep{Govoni2019} respectively.
The spectral index distributions and their relative uncertainties are shown in Figure~\ref{fig:alphaA399_140}.

The spectral index of the radio halo in A\,339 between 140~MHz and 346~MHz appears in the $1.5 \leq \alpha \leq 2$ range and its distribution is rather smooth, although some of the uniformity may be due to limited angular resolution that would average out small scale variations. The spectral index can be assumed uniform across the halo within the uncertainties. The spectral index integrated over the whole radio halo emission is $\alpha = 1.5 \pm 0.14$, slightly steeper than the 140--1400~MHz spectral index of 1.3 assumed by \cite{Govoni2019}, although consistent within the uncertainties, placing it amongst the class of moderately ultra steep radio halos \citep[e.g.][]{Macario2013,Wilber2018}. We scaled the flux density of the halo to 1.4~GHz using $\alpha=1.5$, resulting into a radio power of $P_{1.4} \simeq 1.14 \times 10^{23}$~W~Hz$^{-1}$, well below the correlation between radio halo power and cluster mass \citep[e.g.,][]{Cassano2013,Cuciti21b,Duchesne21} where ultra steep spectrum radio halos are indeed expected \citep{Cassano2013}.


\subsection{Bridge Emission at 346~MHz}
\label{subsec:bridge_emission}


As mentioned in the introduction (Section~\ref{sec:introduction}), \citet{Govoni2019} detected a bridge of radio emission connecting A\,339 and A\,401 with a 
$\sim 1.4$~mJy~arcmin$^2$ average surface brightness at 140~MHz.
The presence of radio emission produced by relativistic particles on scales of a few Mpc poses a question on their acceleration mechanism. 
Due to synchrotron and inverse Compton losses, the particle life time at 140~MHz is of the order of $10^8$~years \citep{Govoni2019}. Therefore particles can only travel one tenth of the bridge extension in their life time, requiring a mechanism of {\it in situ} particle acceleration. \citet{Govoni2019} investigated diffuse shock acceleration, normally considered responsible for radio emission from cluster relics. They found that shock acceleration of thermal electrons would not be sufficiently efficient to achieve the observed emissivity levels. However, re-acceleration of a pre-existing population of mildly relativistic electrons by a population of weak shocks (Mach number of orders of $2-3$) that fills the bridge volume may be able to produce the observed radio emission \citep{Wittor17}. In this case, the spectral index of the bridge would be $\alpha \sim 1.2-1.3$, similar to that observed in relics.

\begin{figure}
	\centering
	\includegraphics[width=1\columnwidth]{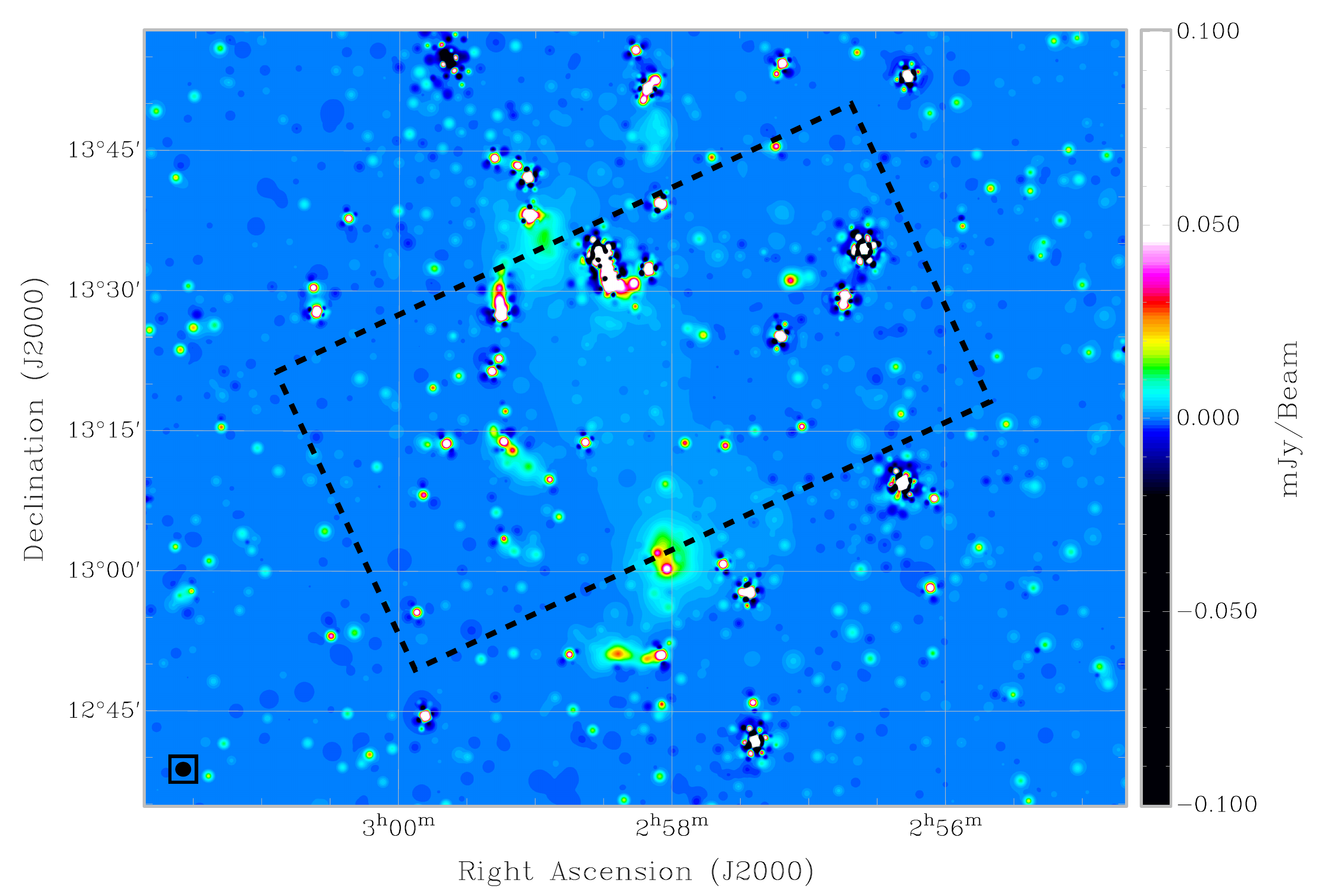}
	\caption{The model image from \citet{Govoni2019}. The region within the black dashed lines highlights the
			model described in Section~\ref{subsec:bridge_emission}. The beam size for the 140~MHz observations ( $72''\times 72''$)  is shown in the inset.}
	\label{fig:lofar_model}
\end{figure}

\begin{figure}
	\centering
	\includegraphics[width=0.95\columnwidth]{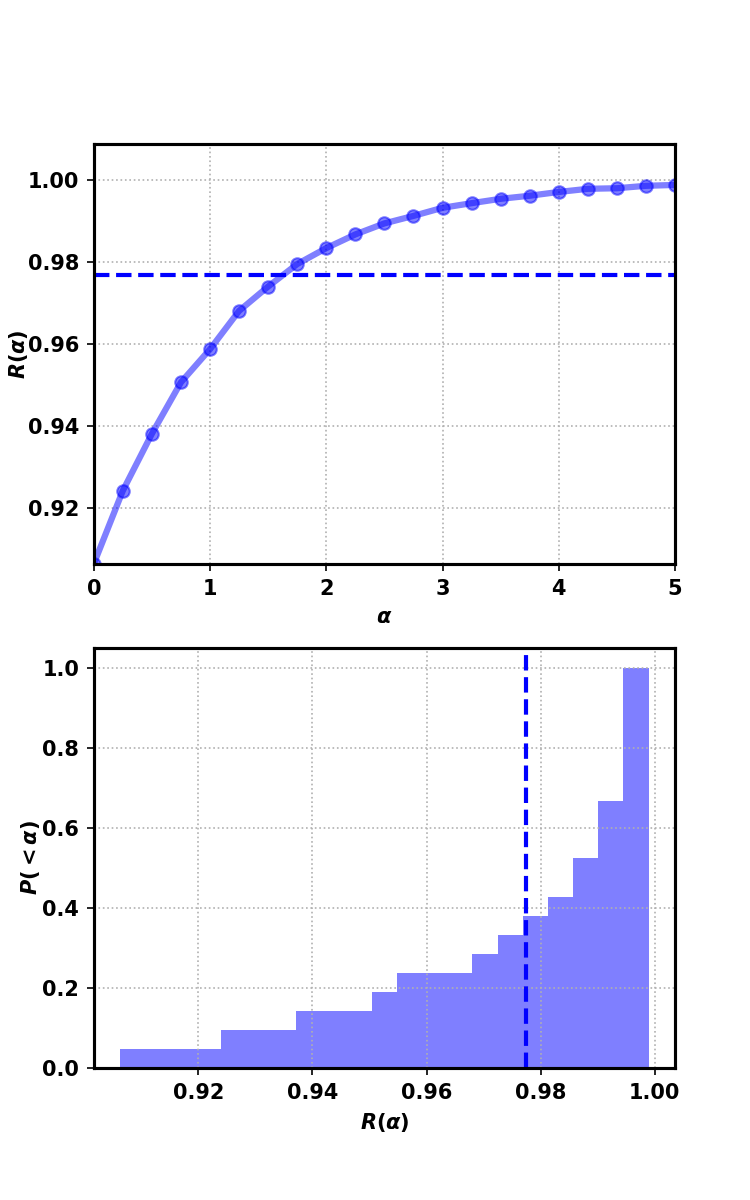}
	\caption{{\textit Upper: }The ratio $R(\alpha)$ described in equation~\ref{eq:ratio_def} as a function $\alpha$. {\textit Lower: }The cumulative distribution of $R(\alpha)$. The dashed lines represent the values of $R(\alpha)$ obtained at $95\%$ confidence interval, corresponding to $\alpha_r = 1.5$.}
	\label{fig:R_alpha}
\end{figure}  

Our 346~MHz observations offer the opportunity to constrain the bridge spectral index and, potentially, discriminate between particle re-acceleration mechanisms. The comparison between the 140~MHz and the 346~MHz images (Figure~\ref{fig:radio_ridge}), shows no evidence of the bridge emission at 346~MHz. As the bridge is an extended, low surface brightness region, a simple rms analysis, usually performed on point sources, would not be a reliable estimate on its flux density. We therefore adopted a procedure similar to the one used to quantify upper limits on the radio halo flux density in cluster observations \citep[e.g.,][]{Venturi2007,Kale2015,Bernardi16,Cuciti21a} in order to set an upper limit on the bridge flux density at 346~MHz, which, conversely, turns into a lower limit on its spectral index. The procedure is as follows:

\begin{enumerate}
	\item we first selected the model image of the bridge emission from the LOFAR image at $\nu_1 = 140$~MHz shown in Figure~\ref{fig:lofar_model}. It consists of the emission above the 3$\sigma$ contour in a $2 \times 3$~Mpc box centered on the bridge - essentially the same region defined in \citet{Govoni2019};
	\item the bridge model image was extrapolated to the central frequencies of the WSRT spectral windows, corresponding to 368, 359, 350, 341, 333, 324, 315~MHz using:
	\begin{equation}
		S_{\nu_2} (x,y, \alpha) = S_{\nu_1} (x,y) \left( \frac{\nu_2}{\nu_1} \right)^{-\alpha},
	\end{equation}
	where $S_{\nu}$ is the flux density of the model image at frequency $\nu$ and position $(x,y)$ and $\alpha$ is the given spectral index. The extrapolated model images were attenuated using the WSRT primary beam model \citep[e.g.,][]{Bernardi2010};
	\item The primary beam-attenuated model images were Fourier transformed into visibilities, added to the calibrated visibilities, then imaged following the same procedure described in Section~\ref{sec:observations}. This procedure takes care of the proper sampling of the bridge emission by the WSRT $uv$-coverage. 
	
	\item We then evaluated the ratio $R(\alpha)$:
	\begin{equation}
		R(\alpha) = \frac{\sum_{x,y=1}^N \tilde{S}(x,y)}{\sum_{x,y=1}^N \big(\tilde{S}(x,y) + S_{\nu_2} (x,y)\big)},
		\label{eq:ratio_def}    
	\end{equation}
	where $\tilde{S}$ is the 346~MHz image (Figure~\ref{fig:radio_ridge}) and $N$ is the total number
	of pixels in the selected region from the model image illustrated in Figure~\ref{fig:lofar_model}. The numerator of equation~\ref{eq:ratio_def} is essentially the flux density calculated over the bridge area from the 346~MHz image and the denominator is the flux density of the 346~MHz image after the bridge model was added to (``injected" into) the visibilities. The ratio $R(\alpha)$ is a monotonically increasing function with $\alpha$:  $\lim_{\alpha \to \infty} R(\alpha) = 1$, implying that the ratio is unity in the limit of no bridge emission.  Conversely, $R(\alpha)$ has a minimum smaller than unity for $\alpha = 0$. This means that there must exist a spectral index value $\alpha_r$ for which $R(\alpha_r)$ is significantly smaller than one, i.e. the injected halo should be detectable above the noise of the 346~MHz observations. 
	
\end{enumerate}
The above procedure is repeated for $0 < \alpha < 5$, with $\Delta \alpha = 0.25$ steps. We then constructed the cumulative distribution of $R(\alpha)$ normalized to unit area over the chosen interval. The results are plotted in Figure~\ref{fig:R_alpha}.
	The upper panel of Figure~\ref{fig:R_alpha} demonstrates the aforementioned monotonic behaviour of $R(\alpha)$ as $\alpha$ increases, while the bottom panel shows its corresponding cumulative distribution. In order to determine the previously defined $\alpha_r$, we calculated the 95$\%$ confidence interval from the cumulative distribution. 
	Resulting  bounds are indicated by the blue dashed lines in Figure~\ref{fig:R_alpha} and this value corresponds to $\alpha=1.5$. 
	We could therefore say that $P(< \alpha) = 95\%$ for $\alpha_r = 1.5$, in other words, the probability of finding an excess flux density with respect to the 346~MHz observations would be 95\% if the spectral index of the bridge was smaller than 1.5. Given no detection at 346~MHz, this result sets a lower limit on the spectral index of the bridge $\alpha > 1.5$ at 95\% confidence level.

\begin{figure*}
	\centering
	\includegraphics[width=2.1\columnwidth]{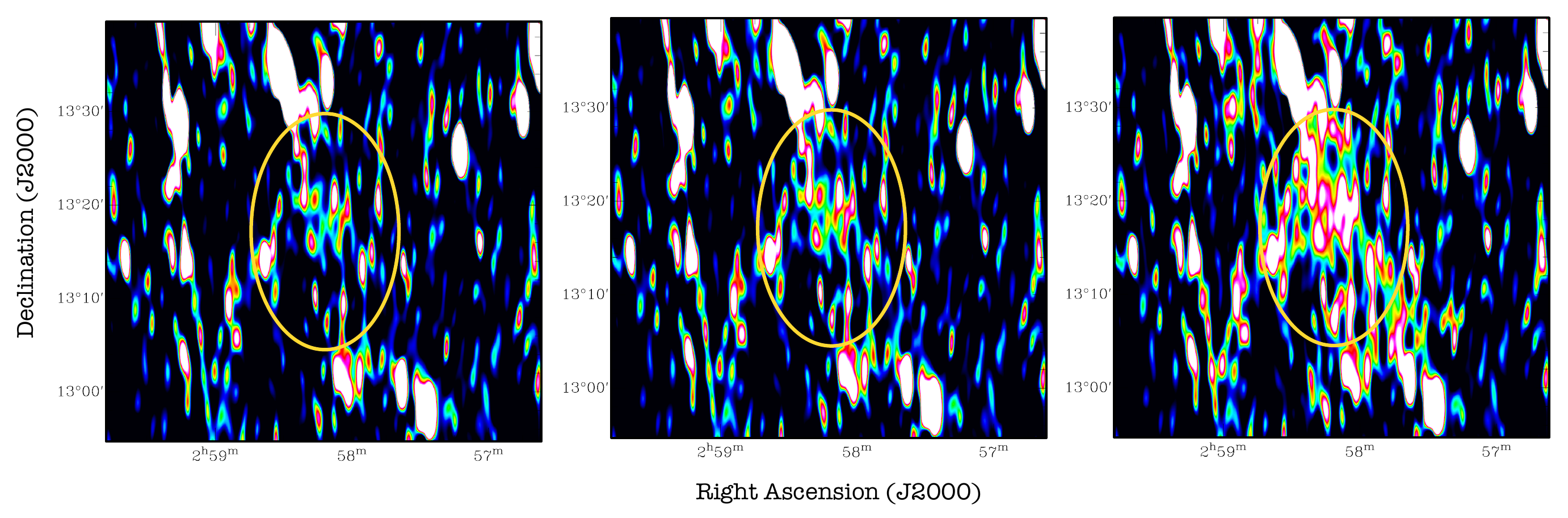}
	\caption{{\textit Left:} WSRT observations at 346~MHz (same as Figure~\ref{fig:radio_ridge}, but centred around the bridge emission. {\textit Middle:} WSRT observation imaged after injecting the bridge emission at $\alpha=1.5$. {\textit Right:} WSRT observation imaged after injecting the bridge emission at $\alpha=0$. The yellow circle highlights the region where the bridge emission is most prominent. The synthesized beam is $205'' \times 43''$.}
	\label{fig:injected_images}
\end{figure*}

The detection of the bridge emission in our observations at $\alpha=1.5$ is 
illustrated in Figure~\ref{fig:injected_images}. The region within the yellow circle highlights the area where the change in $R(\alpha)$ is expected. We show images without the mock bridge emission (left panel), and with the injected bridge emission at $\alpha=1.5$ (middle panel) and $\alpha=0$ (right panel). We clearly see the rise in flux density at $\alpha=1.5$, contributing to the change in $R(\alpha)$ demonstrated in Figure~\ref{fig:R_alpha}. At $\alpha=0$, we are essentially looking at the full bridge emission at 140~MHz from \citep{Govoni2019}, justifying the steep gradients in $R(\alpha)$ between $0 < \alpha < 1.5$.

It should be noted that the method adopted to estimate this constraint may be affected by the difference in $uv$ coverage from the different arrays as well as the presence of bright sources. \citet{Govoni2019} could not reliably evaluate the spectral index on the bridge emission using the Very Large Array observations at 1.4~GHz \citep{Murgia2010} because of the difference in baselines. Our almost filled-$uv$ coverage (Figure~\ref{fig:uvcov_92cm}) provides us with a comparable angular resolution to \citet{Govoni2019}, with a drop of $\sim$~2 on one side of the synthesized beam. Additionally, we masked sources above 3$\sigma$ as described in (i), thus, mitigating the effect of nearby bright sources.

\subsection{Relic candidates at 346~MHz}
\label{subsec:relics}

\begin{figure*}
\centering
\includegraphics[width=2.0\columnwidth]{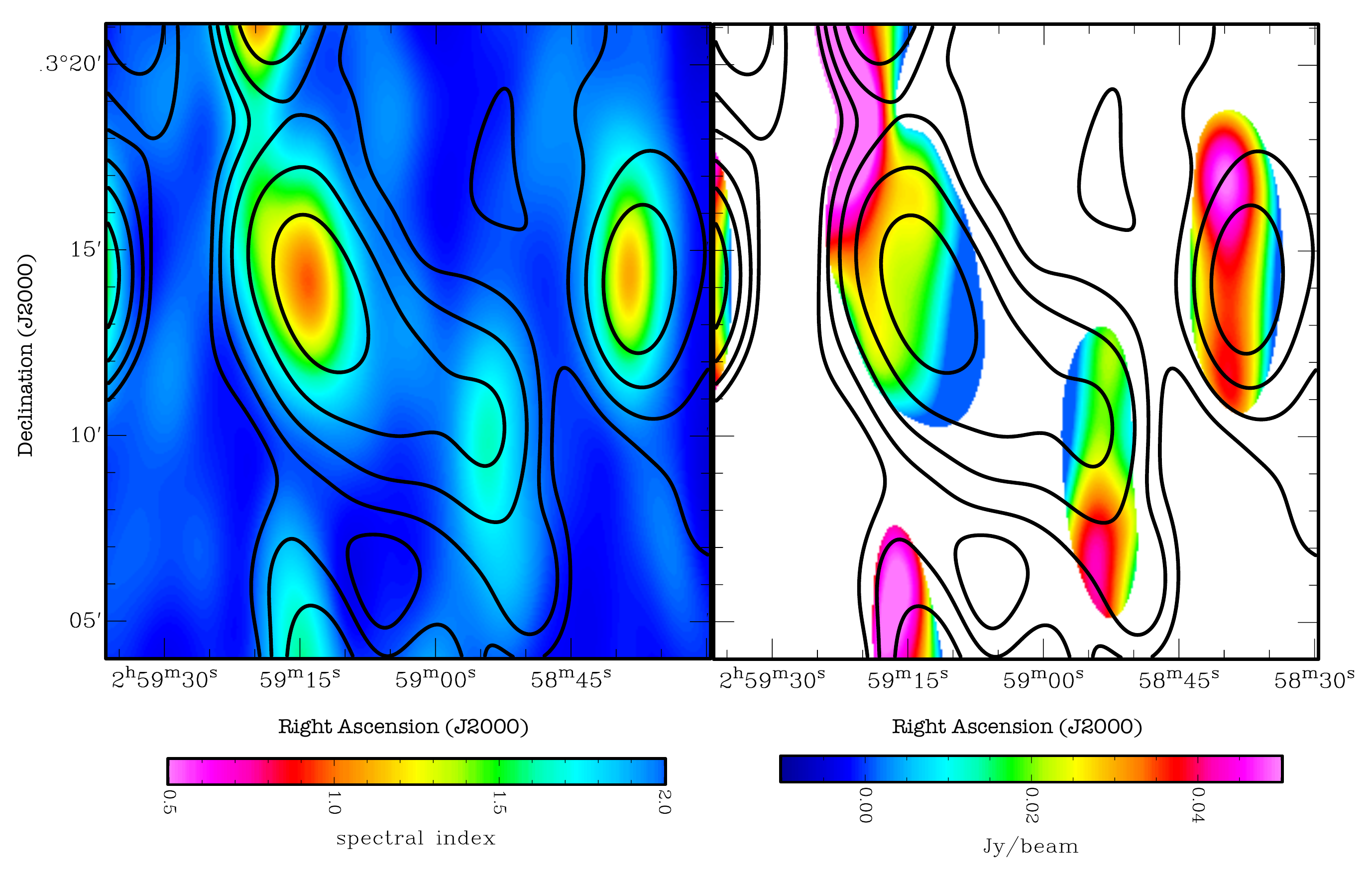}
\caption{Left: The 346~MHz image overlaid with 140~MHz contours of the diffuse source $d$ \citep{Govoni2019}. Both images were smoothed to the same angular resolution of $205'' \times 72''$ (same as Figure~\ref{fig:radio_ridge}). The first contour is drawn at 4~mJy~beam$^{-1}$ (3$\sigma$, with an rms noise of 1.3~mJy~beam$^{-1}$), and the following ones are spaced by a factor of two. The surface brightness peaks match very well between the two observing frequencies, although the interconnecting region is not very visible at 346~MHz. Right: Same contours as left, but overlaid on the radio spectral index map of sources $d$ between 140~MHz and 346~MHz. We note a steepening of the spectral index across the source. Blank regions correspond to pixels fainter than $5\sigma$ at either 140~MHz or 346~MHz.}
\label{fig:relicd_140MHz}
\end{figure*}

\begin{figure*}
\centering
\includegraphics[width=2.0\columnwidth]{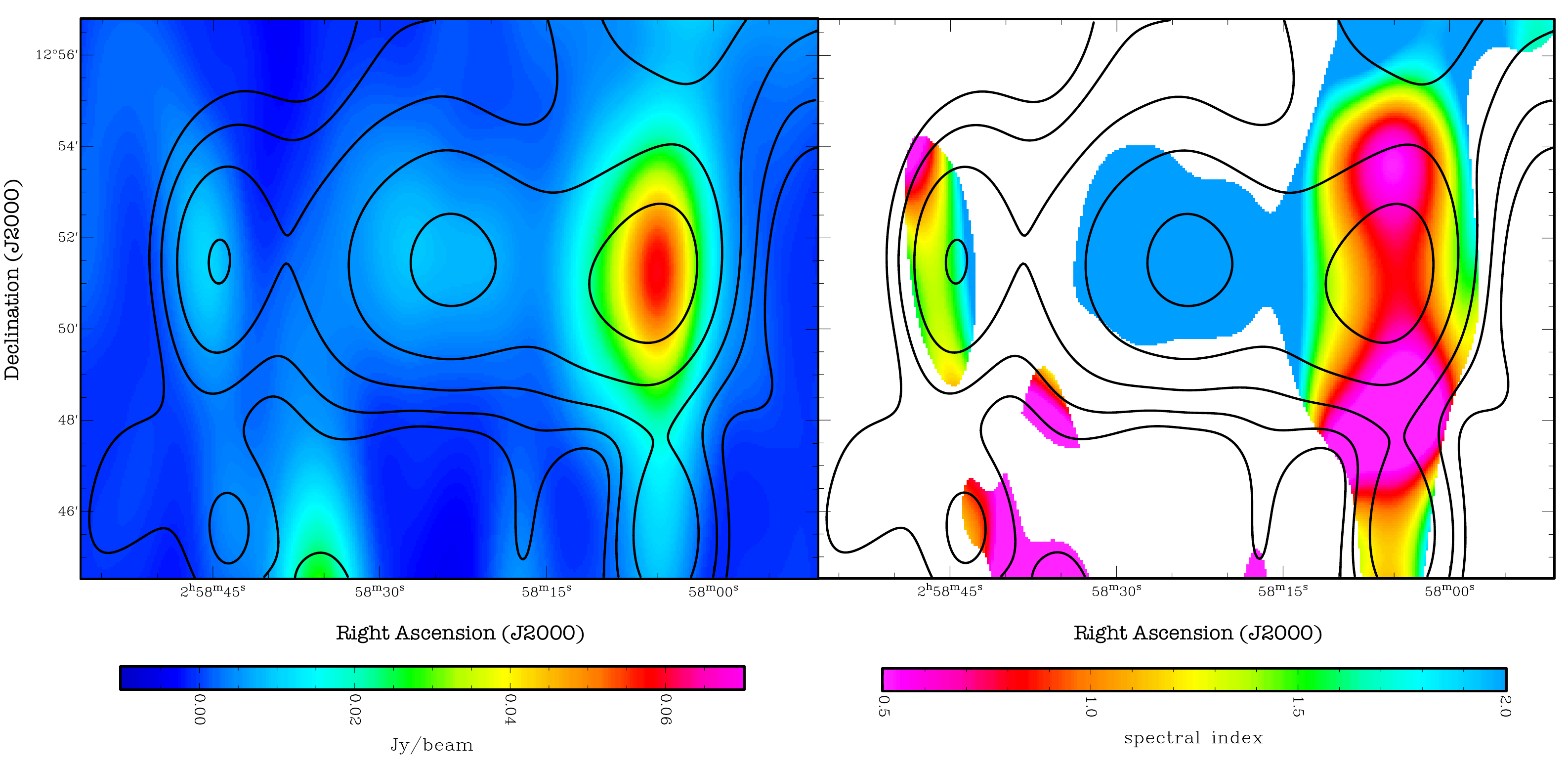}
\caption{Same as Figure~\ref{fig:relicd_140MHz} but for source {\it f}. As for source {\it d}, the spectral index is steeper in low surface brightness regions.}
\label{fig:relicf_140MHz}
\end{figure*}

\citet{Govoni2019} also detected two diffuse sources - $d$ and $f$ in Figure~\ref{fig:radio_ridge} - with no obvious optical counterpart and that could not be unambiguously classified. Their high resolution (10~arcsec) observations show head-tail morphologies for both sources, indicating that they may be faint radio galaxies with switched-off tails.
The presence of an X-ray shock \citep{Akamatsu17} in the proximity of source $d$, would support the relic hypothesis.
Both sources are visible at 346~MHz, with a similar morphology compared to the 140~MHz (Figures~\ref{fig:relicd_140MHz} and \ref{fig:relicf_140MHz}). Source $d$ is somewhat resolved into two brightness peaks at 346~MHz, and the interconnecting region appears to be at noise level.  

The spectral index distribution has a trend across source $d$, with steep values ($\alpha \sim 1.2 - 1.3$) corresponding to the brightness distribution peaks, and values that become ultra-steep ($\alpha > 2$) in the interconnecting region. In the case of source $f$, the core emission shows a flatter spectral index ($\alpha \sim 0.5 - 0.8$) which becomes steeper ($\alpha > 1.5$) across the source. Results for both sources are consistent with the spectral index distributions derived between 140~MHz and 1.4~GHz in \cite{Govoni2019}, albeit at higher angular resolution (10~arcsec).

The integrated spectral indices of relics {\it d} and {\it f} were derived by integrating the brightness distribution at 346~MHz above the $5\sigma$ contours used in \citep{Govoni2019}. We obtained flux densities $S = 90 \pm 5$~mJy and $S = 141 \pm 7$~mJy for sources {\it d} and {\it f} respectively, yielding steep spectral indices $\alpha = 1.10 \pm 0.14$ (source {\it d}) and $\alpha = 1.46 \pm 0.14$ (source {\it f}). Integrated spectral indices are also consistent with \cite{Govoni2019}, 1.23 and 0.96 for sources {\it d} and {\it f }respectively,  albeit a bit steeper for source $f$, likely due to the fact that our observations are more sensitive to the steep spectrum region compared to the 1.4~GHz data used in \cite{Govoni2019}.




\section{Discussion and Conclusions}
\label{sec:conclusions}

We presented radio observations of the A\,399--A\,401 cluster pair at 1.7, 1.4, 1.2~GHz and 346~MHz with the WSRT telescope, focusing on the analysis of the diffuse radio emission. 
This cluster pair is a rare case: it appears to be a system in its early merging state, connected by a 3~Mpc-long mass filament with a $4.3 \times 10^{-4}$~cm$^{-3}$ density \citep{Govoni2019}. Both clusters host a radio halo, detected at 1.4~GHz \citep{Murgia2010} and 140~MHz \citep{Govoni2019}. Observations at 140~MHz also reveal the presence of a low surface brightness radio bridge connecting both clusters as well as two other diffuse features that may be either switched-off, tailed radio galaxies or relics.

Our observations at GHz frequencies were not sufficiently deep to improve the existing radio halo images at 1.4~GHz. At 346~MHz, we detected the A\,399 radio halo at high significance, with a $125 \pm 6$~mJy flux density. The halo has a similar round-shaped morphology to the 140~MHz observations, but its size is comparatively smaller.

We then searched our 346~MHz observations for the presence of the 3~Mpc-extended synchrotron emission bridge that connects both halos observed at 140~MHz along with two diffuse emission features with no obvious optical counterparts observed in \citep{Govoni2019}.
Both diffuse sources are visible at 346~MHz and their spectral indices between 346~MHz and 140~MHz are $\alpha = 1.1 \pm 0.14$ (source {\it d}) and $\alpha = 1.46 \pm 0.14$ (source {\it f}). 
The values of the integrated spectral indices together with the spectral index steepening across the source - in both cases - suggest that they are more likely switched off radio galaxies rather then relics, or relic emission connected with a radio galaxy \citep[e.g.][]{Bonafede14}.

The bridge was not visible in the 346~MHz data and we, therefore, used them together with the 140~MHz observations to set a constraint on its average spectral index through the ``injection" method. We found a lower limit on the spectral index to be $\alpha > 1.5$ at 95\% confidence level. Such a steep spectral index value cannot be easily explained by diffusive shock acceleration even if an initial population of mildly relativistic electrons is assumed \citep{Govoni2019}. The result is instead more aligned with the expectations of second--order Fermi mechanisms, where particles are accelerated and magnetic fields amplified by turbulence \citep{Brunetti2020}. 

\section*{acknowledgements}
This research is partly supported by the Australian
Research Council Centre of Excellence for All Sky Astrophysics in 3 Dimensions (ASTRO 3D), through project number CE170100013.
This work is also supported by the National Research Foundation under grant 92725. Any opinion, finding and conclusion or recommendation expressed in this material is that of the author(s) and the NRF does not accept any liability in this regard. We acknowledge the use of the \citet{Wright06} cosmology calculator. This research is supported by the South African Research Chairs Initiative of the Department of Science and Technology and National Research Foundation. VV and MM acknowledge support from INAF mainstream project``Galaxy Clusters Science with LOFAR” 1.05.01.86.05. GB acknowledges helpful discussions with Gianfranco Brunetti and Franco Vazza on particle acceleration mechanisms.

\section*{Data Availability}

The data presented in this paper and the software used can be shared upon reasonable request to the corresponding author.

\bibliographystyle{mnras}
\bibliography{bibliography}

\appendix
\section{Results at high frequencies (GHZ)}
\label{appendix:GHzimages}
\begin{figure}
\centering
\includegraphics[width=1.05\columnwidth]{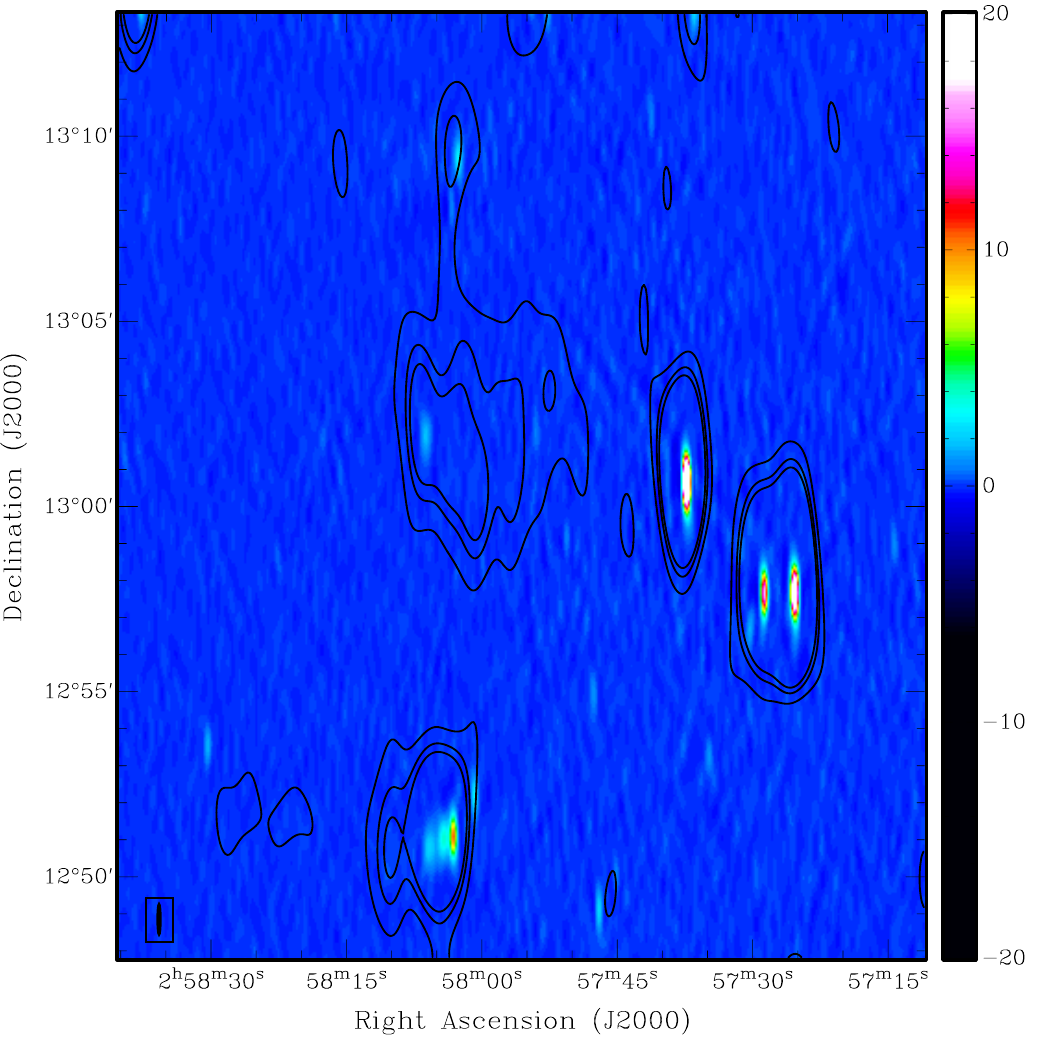}
\caption{The 1.7~GHz image, centred on the A\,399 cluster, overlaid with 346~MHz contours. The first contours are drawn at 4$\sigma$, where $\sigma$ is the noise rms reported in Table~\ref{tab:table_obs}, and then spaced by a factor of two. The radio halo is clearly visible at 346~MHz but no corresponding emission is detected at 1.7~GHz. The beam size for the 1.7~GHz observations ( $53''\times 9''$)  is shown in the inset and intensity displayed on the colorbar is in mJy/Beam.}
\label{fig:1_7GHz}
\end{figure}

\begin{figure}
\centering
\includegraphics[width=1.05\columnwidth]{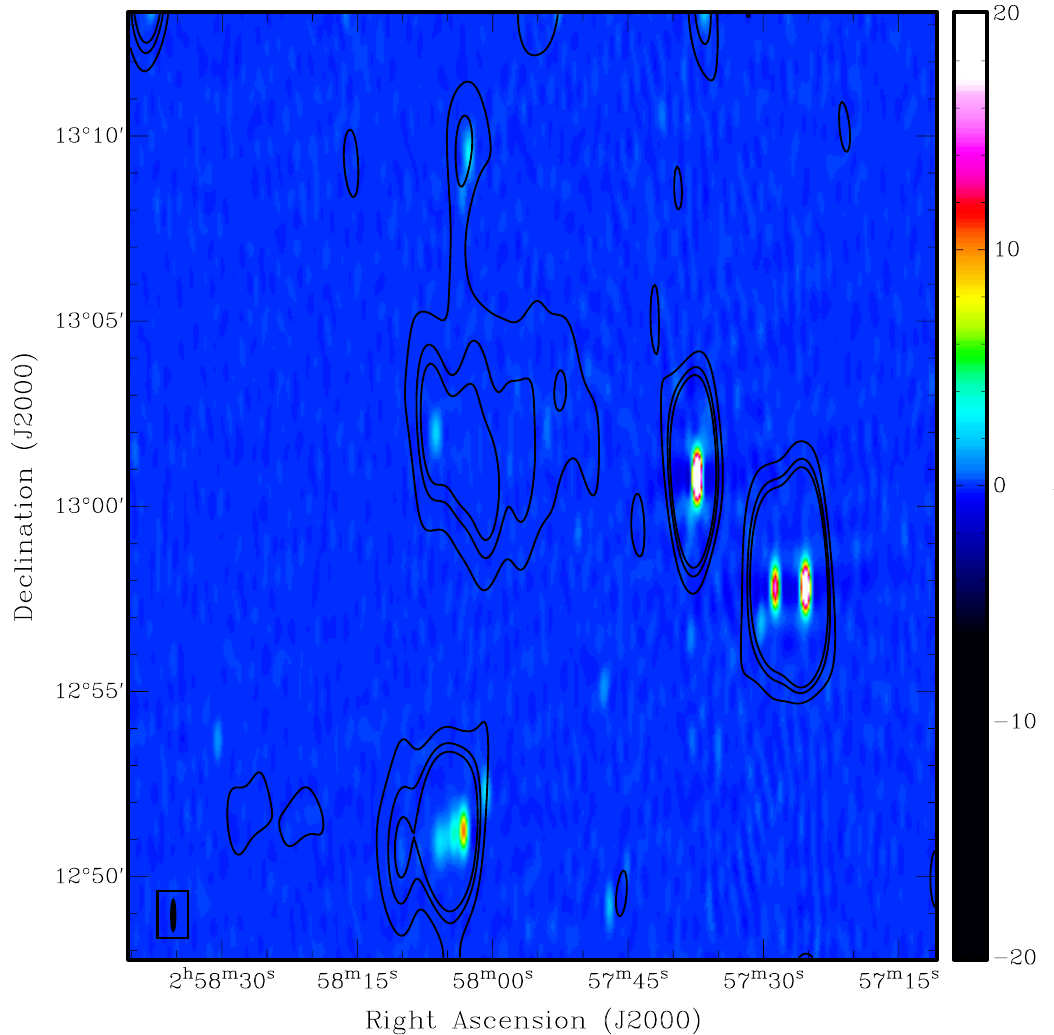}
\caption{Same as Figure~\ref{fig:1_7GHz} but for the  1.4~GHz observations.}
\label{fig:1_4GHz}
\end{figure}

\begin{figure}
\centering
\includegraphics[width=1.05\columnwidth]{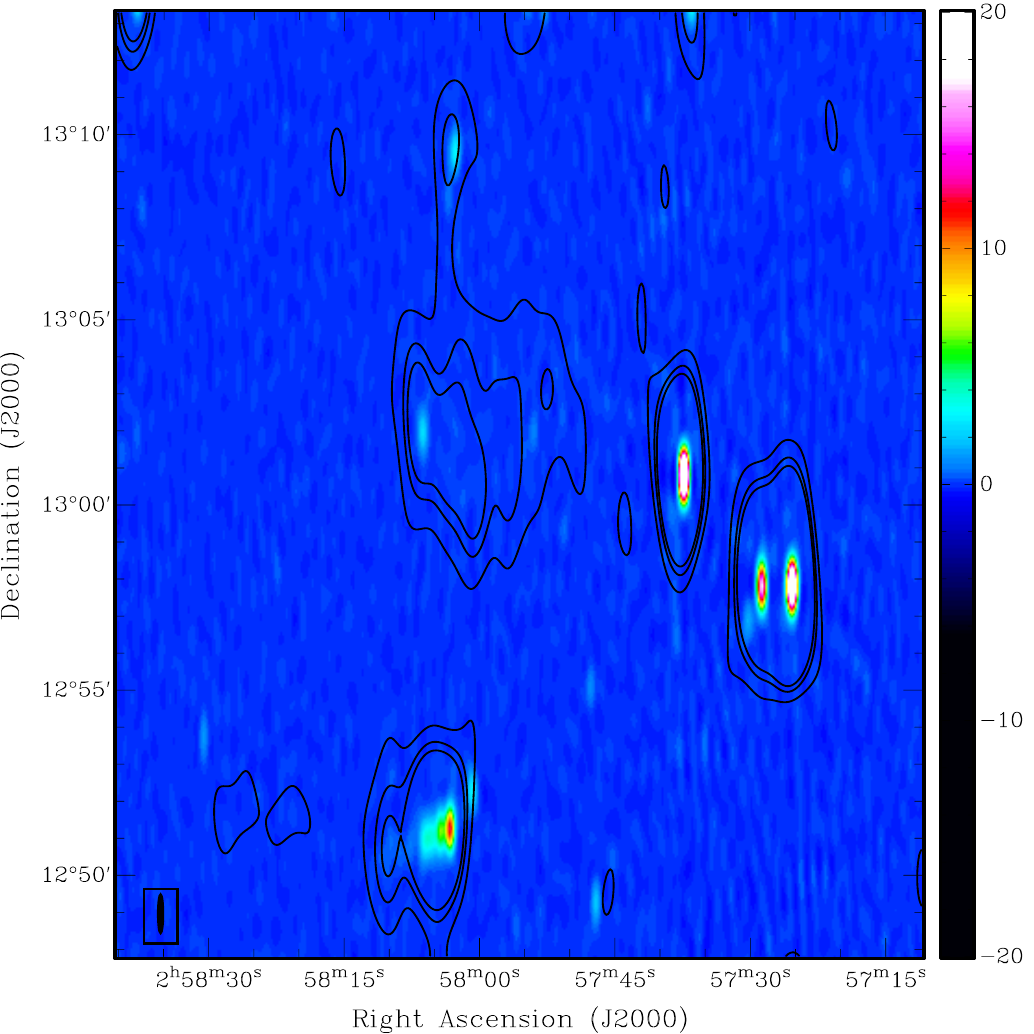}
\caption{Same as Figure~\ref{fig:1_7GHz} but for the  1.2~GHz observations.}
\label{fig:1_2GHz}
\end{figure}

As discusses in Section~\ref{sec:results}, the WSRT observations at 1.2, 1.4 and 1.7~GHz do not show clear detection of the A\,399 radio halo. We present images at these aforementioned frequencies in Figures~\ref{fig:1_7GHz}, \ref{fig:1_4GHz} and \ref{fig:1_2GHz}.

\bsp	
\label{lastpage}
\end{document}